\documentclass[11pt,a4paper]{article}
\usepackage{jheppub}
\usepackage{bm}
\usepackage{feynmf}
\usepackage{mathrsfs}
\usepackage{slashed}
\usepackage{mathtools}
\usepackage[dvipsnames]{xcolor}
\usepackage{physics}
\usepackage{hyperref}
\usepackage{graphicx, animate}
\usepackage{tikz}
\usepackage{listings}
\usepackage{orcidlink}

\usepackage[utf8]{inputenc}

\usepackage{xcolor}
\usepackage{parskip}

\usepackage{soul}

\usepackage[outline]{contour}
\contourlength{2pt}
\usepackage{amsfonts}

\usepackage[mode=buildnew]{standalone}

\allowdisplaybreaks
 
\usetikzlibrary{3d} 
\usetikzlibrary{shadows.blur}
\usetikzlibrary{fadings}
\usetikzlibrary{decorations.text} 
\usetikzlibrary{patterns}

\colorlet{mydarkblue}{blue!40!black}
\colorlet{mylightblue}{mydarkblue!12} 
\colorlet{myred}{red!80!black}
\colorlet{mydarkred}{red!50!black}
\colorlet{mylightred}{mydarkred!12}
\colorlet{mydarkgreen}{green!30!black}
\colorlet{mylightgreen}{mydarkgreen!12}
\colorlet{myorange}{orange!63!black}
\colorlet{mylightorange}{orange!80!black!12}

\usetikzlibrary{patterns} 
\def\hatchsize{4pt}
\makeatletter
\pgfdeclarepatternformonly{myhatch}{%
  \pgfqpoint{-1pt}{-1pt}}{\pgfqpoint{\hatchsize}{\hatchsize}}{\pgfqpoint{\hatchsize}{\hatchsize}
}{%
  \pgfsetlinewidth{0.3pt}
  \pgfpathmoveto{\pgfqpoint{0pt}{\hatchsize}}
  \pgfpathlineto{\pgfqpoint{\hatchsize}{0pt}}
  \pgfusepath{stroke}
}
\makeatother

\tikzfading[
  name=fade out,
  inner color=transparent!0,
  outer color=transparent!100
]
\tikzfading[
  name=fade in,
  inner color=transparent!100,
  outer color=transparent!0
]
\tikzfading[
  name=fade right,
  left color=transparent!0,
  right color=transparent!100
]

\usetikzlibrary{shadings}
\makeatletter
\pgfdeclareradialshading[tikz@radial@inner,tikz@radial@outer]%
  {myradial}{\pgfpointorigin}{
    color(0bp)=(pgftransparent!100);%
    color(10bp)=(pgftransparent!90);%
    color(50bp)=(pgftransparent!0)%
  }
\makeatother
\pgfdeclarefading{myradial}{\pgfuseshading{myradial}}%

\tikzset{pics/.cd,
    arrow1/.style,
    code={
    \draw[-latex] (0,0) to [out=180,in=235] (5,-5);
}}

\graphicspath{ {./main_images/} }

\title{Do we live on the End of the World?}

\author{Benjamin Muntz \orcidlink{0000-0002-0183-8783},}
\author{Antonio Padilla \orcidlink{0000-0002-2672-0589},}
\author{Paul M. Saffin \orcidlink{0000-0002-4290-3377}}
\date{~}

\affiliation{School of Physics and Astronomy, University of Nottingham, University Park, Nottingham NG7
2RD, United Kingdom}

\affiliation{Nottingham Centre of Gravity, University of Nottingham, University Park, Nottingham NG7 2RD, United Kingdom}

\emailAdd{benjamin.muntz@nottingham.ac.uk}
\emailAdd{antonio.padilla@nottingham.ac.uk}
\emailAdd{paul.saffin@nottingham.ac.uk}

\abstract{
\noindent We propose a scenario of a de Sitter universe living on an End-of-the-World brane. Motivated by the Swampland programme and in particular the Cobordism Conjecture, we consider a compact region of AdS$_5$ nucleating from nothing, with a dS$_4$ living on its boundary. We show that it can equivalently be interpreted as an up-tunnelling from AdS$_5$ with cosmological constant $\Lambda\to - \infty$, following Brown and Dahlen's proposal for `nothing'. Their picture na\"ively leads to the conclusion that the brane has infinite negative tension. But we show that it becomes finite and positive once we employ holographic renormalization, recovering the Bubble of Something where the domain wall becomes a boundary of spacetime. The same holds true in any number of dimensions and, moreover, at the level of metric perturbations. This provides motivation for alternative routes of obtaining cosmology from quantum gravity or string theory using domain walls, departing from conventional vacuum approaches.
}

\begin{document} 
\maketitle
\flushbottom

\section{Introduction}
A wealth of observational evidence points to a universe that is undergoing an accelerated expansion \cite{Planck:2018vyg,SupernovaCosmologyProject:1998vns,SupernovaSearchTeam:1998fmf}, consistent with a late time de Sitter (dS) vacuum. However, constructing dS vacua from the top down remains a great challenge in modern physics. Despite many interesting attempts to obtain them directly from fundamental theory (for example \cite{Kachru:2003aw, Louis:2012nb, Cicoli:2013cha,Blaback:2013ht, Antoniadis:2019rkh,Cribiori:2019hrb,Crino:2020qwk}), there is little doubt that they are hard to find and even harder to do so consistently, with full perturbative control. Indeed, the path towards finding dS vacua in string theory must get round a number of no-go theorems  \cite{Maldacena:2000mw, Kutasov:2015eba, Green:2011cn} and is paved with a range of potential pitfalls. For a sample of some of the things that can go wrong, see  \cite{Moritz:2017xto,Gautason:2013zw,Bena:2009xk,Danielsson:2014yga, Michel:2014lva, Liu:2024sgb}. As a result of these difficulties, it has been conjectured that dS vacua may even be fundamentally incompatible with, at the very least, string theory \cite{Obied:2018sgi, Ooguri:2018wrx, Andriot:2018wzk, Garg:2018reu}. Or in other words, they lie in the \emph{Swampland} of gravitational effective field theories. Although this view is not without nuance and opposition (see, for example, \cite{Hebecker:2018vxz, Akrami:2018ylq, Casse:2022ymj}), it is safe to say that the debate surrounding dS remains both lively and multi-faceted; for recent reviews on this topic, we refer the reader to \cite{Danielsson:2018ztv, Dine:2020vmr, Berglund:2022qsb, VanRiet:2023pnx}. Obtaining a model of quintessence in string theory, where the late time acceleration is due to a dynamical scalar field in slow roll, is arguably even more challenging \cite{Cicoli:2018kdo, ValeixoBento:2020ujr, Hebecker:2019csg,Cicoli:2021fsd, Cicoli:2021skd,Hebecker:2023qke, Cicoli:2024yqh}. 

In this work, we will attack the problem of obtaining dS slightly differently to the mainstream approach found in the literature. Typically, you take your favourite string theory, compactify on some Calabi-Yau, add necessary ingredients such as fluxes and branes, and then check that the resulting effective theory is robust against $g_s$ and $\alpha'$ corrections. Whilst it goes without saying that heroic efforts have been made in this direction, such a strategy also involves a preliminary choice: that we expect our universe to truly represent a vacuum state of the full theory. If the Landscape is vast, this is by no means inconceivable. But by making this choice, we omit by default a separate possibility that could perhaps be as numerous as (if not more than) the number of vacua: transitions between them. To emphasise, if some moduli locally tunnel between different vacuum configurations, it gives rise to domain walls partitioning these spatially separated regions. The question then becomes whether we can straightforwardly put dS on one of these domain walls. We advocate that the answer to this question is yes.

Local physics on a domain wall is necessarily sensitive to the spacetime regions that it delimits and typically demands a certain hierarchy of scales between them in order to recover any realistic cosmology. One of our goals with this work is to skip this additional requirement by focusing on a very particular process: the decay of nothing into something. In other words, we consider the absence of spacetime and its content as the initial state and allow it to nucleate a bubble of spacetime. Although bizarre at first glance, it is instructive to first discuss the inverse process where spacetime decays to nothing, also known as the ``Bubble of Nothing''. In the earliest version by Witten \cite{Witten:1981gj}, he demonstrates that the Kaluza-Klein vacuum admits an instability that allows it to decay via bubble nucleation. As one approaches the bubble from the outside, the size of the internal $S^1$ dynamically shrinks until it reaches zero size and `pinches off'. This zero-size locus, called an End-of-the-World (ETW) brane, becomes a new boundary of the spacetime in the sense that any point inside the bubble is infinite geodesic distance away. A way to understand this is by recognising that the lower-dimensional gravitational coupling diverges, $\kappa_4^{-2}\sim \mathrm{Vol}(S^1)\kappa_5^{-2}\to 0$. In other words, there is effectively no spacetime inside the Bubble of Nothing, as the name suggests. The reversed process described by the same instanton, dubbed the ``Bubble of Something'', should describe the creation of a spacetime from nothing. Alongside Bubbles of Nothing and ETW branes, they have gained a lot of traction within the Swampland programme in recent years \cite{Friedrich:2023tid, Angius:2023uqk, Blumenhagen:2023abk, Huertas:2023syg, Blanco-Pillado:2023aom, Blanco-Pillado:2023hog, Delgado:2023uqk}, seeded by the Cobordism Conjecture \cite{McNamara:2019rup}. Mathematically, cobordism is an equivalence relation between compact manifolds; two manifolds of the same dimension, say $\mathcal{M}$ and $\mathcal{M}'$, are said to be cobordant if there exists a manifold of one dimension higher, $\mathcal{W}$, whose boundary is the disjoint union of the two, $\partial \mathcal{W}=\mathcal{M}\sqcup \mathcal{M}'$. Taking Witten's Bubble of Nothing as an example, the pinching off of the $S^1$ can be seen as a cobordism to nothing, since $S^1$ is itself a boundary. The Cobordism Conjecture then states that, in quantum gravity, the group consisting of all such equivalence classes contains only the trivial element, i.e. any compactification can be related to nothing by a cobordism. In that regard, the conjecture predicts that the associated ETW branes \emph{must} be included in the spectrum of domain walls. If that is the case, it appears natural for us to consider whether they can also serve as braneworlds.  This will be the focus of our work. 

An interesting proposal was made by Brown and Dahlen that `nothing' can in fact be thought about as Anti-de Sitter (AdS) space with $\Lambda\to -\infty$, where $\Lambda$ is the cosmological constant \cite{Brown:2011gt}. Or equivalently, take the AdS length scale to zero, $\ell\to 0$. It is appealing to ask whether this interpretation resonates with ideas from the Swampland programme. The subtlety in taking this limit, however, is that it by design leads to diverging quantities, potentially spoiling the effective description. For example, the brane tension will turn out to be negative for up-tunnelling processes, and we will show how it diverges in the aforementioned limit. The same was pointed out in \cite{Friedrich:2023tid} where the authors necessitate some renormalization procedure to make the Brown-Dahlen interpretation consistent. The need to renormalize should not be unexpected. After all,  the classical gravity description is not expected to hold in the limit where $\ell\to 0$, and so divergences are expected to occur from pushing the effective field theory beyond its regime of validity.  The main part of this work will be to show precisely how to realise the required renormalization using holography.

Before outlining the content of this paper, allow us to contrast with another similar idea, namely the ``dark bubble cosmology'' \cite{Banerjee:2019fzz, Banerjee:2020wix, Banerjee:2020wov, Banerjee:2021qei, Banerjee:2021yrb, Danielsson:2022fhd, Danielsson:2022odq, Danielsson:2022lsl, Banerjee:2022ree, Basile:2023tvh, Danielsson:2023alz, Banerjee:2023uto, Danielsson:2024frw}. It is important to stress that it is inherently different from what we will discuss -- however, the intuition shares similar features and will be of benefit to keep in mind. The motivation for the dark bubble stems from the non-supersymmetric AdS Instability Conjecture \cite{Ooguri:2016pdq}, which poses that any such AdS geometry supported by fluxes, must possess an instability. Provided this statement, the dark bubble considers the decay $\mathrm{AdS}_5^+\to \mathrm{AdS}_5^-$ (here we use $+$/$-$ to denote the parent/daughter vacua) via the nucleation of a spherical co-dimension one brane. If there is a hierarchy $\Lambda_-<\Lambda_+<0$ between bulk cosmological constants, i.e. it is a down-tunnelling process, the junction conditions admit positive tension bubble walls and finite nucleation rates. Many aspects of the dark bubble have been explored in recent years, including black holes \cite{Banerjee:2021qei, Danielsson:2024frw}, gravitational waves \cite{Danielsson:2022fhd}, electromagnetism \cite{Basile:2023tvh}, and string theory embeddings \cite{Danielsson:2022odq, Danielsson:2022lsl}. 
 
In the spirit of Brown and Dahlen, the Bubble of Something can be viewed as an up-tunnelling event with $\Lambda_+<\Lambda_-<0$, and $\Lambda_+ \to -\infty$. As we have already mentioned above, this na\"ively requires a bubble wall of negative tension to support the transition. However, this is only true for the bare tension -- the properly renormalized tension turns out to be positive. Of course, the Bubble of Something as an up-tunnelling is inherently different from the dark bubble scenario, where down-tunnelling occurs. We would also like to highlight two additional, differing features, which will play a pivotal r\^ole in our discussion. First is the fact that the dark bubble requires the non-normalizable graviton zero mode in order to give the correct effective graviton propagator on the bubble. This can be sourced by radially stretched bulk strings attached to the bubble. For our model, we will argue that the non-normalizable mode is not needed to recover the usual four-dimensional gravity, and that it is nevertheless suppressed in the $\Lambda_+\to -\infty$ limit. Second is the emphasis that the dark bubble is fundamentally different from the Randall-Sundrum models \cite{Randall:1999vf,Kaloper:1999sm}, since there is no $\mathbb{Z}_2$ symmetry across the brane. They are, in fact, asymmetric braneworlds \cite{Padilla:2004mc, Padilla:2004tp, Gergely:2003pn, Gergely:2004ax} with a well-defined notion of ``inside'' and ``outside'' \cite{Banerjee:2022ree}. Another way to see this is that only the outside patch occupied by the initial vacuum state possesses a conformal boundary. Similarly, we will also argue that the Bubble of Something, despite it also admitting an inside and outside patch, can be interpreted as a one-sided version of a Randall-Sundrum like model.

Finally, we would like to emphasise the r\^ole played by Euclidean methods in delivering a dS universe on a bubble wall.  In $D$ dimensions, the dominant instanton is expected to exhibit $O(D)$ symmetry \cite{Coleman:1977th,Oshita:2023pwr}, with the domain wall corresponding to a $(D-1)$-dimensional spherical hypersurface at fixed radius. Upon Wick rotating this solution to Lorentzian signature, we see that the sphere becomes an expanding bubble wall with a $(D-1)$-dimensional dS geometry. 

The plan for the rest of this paper is as follows: in Section \ref{sec:bubbleofsomething} we will compute the Bubble of Something instanton in two ways -- first by interpreting `nothing' as a vanishing contribution to the action, and second by considering it as AdS and subsequently taking the limit $\Lambda\to -\infty$. We discuss multiple interpretations of the instanton and provide the dS solution on the ETW brane. In Section \ref{sec:renormalization} we argue that the infinities associated with the limit are artefacts of an unrenormalized theory. We give a basic introduction to holographic renormalization and use it to show that the different interpretations of Section \ref{sec:bubbleofsomething} are equivalent. We also discuss metric perturbations and the localisation of gravity on the brane in each case. Appendix \ref{sec:appendixA} provides a verification that these results can also be obtained from standard braneworld computations.

\section{Bubbles of Something -- Two Perspectives}\label{sec:bubbleofsomething}
In this section, we consider a dS$_4$ brane as the boundary of a bubble of AdS$_5$. We shall do this in two different ways. First, we consider the Euclidean instanton action where the bubble wall corresponds to an ETW brane cutting off spacetime entirely. From this perspective, `nothing' is simply interpreted as a vanishing contribution to the action. The instanton action is thus rendered finite because the otherwise divergent volume integral is cut off from the conformal boundary. Alternatively, we perform a general computation of $\mathrm{AdS}_5^+\to \mathrm{AdS}_5^-$ vacuum decays. I.e. there is a bubble of $\mathrm{AdS}_5^-$ nucleating inside a parent $\mathrm{AdS}_5^+$, with $\pm$ referring to two different vacuum energies. In that case the instanton action will in general also depend on the AdS length $\ell_+$ of the parent vacuum. We nevertheless confirm that in the limit $\ell_+\to 0$, the bounce yields precisely the same result as in the first case. However, the instanton action will differ by a divergent term that can be interpreted as a tension. Later we show that this term is precisely cancelled by holographic renormalization \cite{Skenderis:2002wp, Papadimitriou:2016yit}. 

\subsection{One-sided AdS bubble nucleation}
We start with the nucleation of a bubble of AdS$_5$ which has some boundary $\Sigma$ carrying a tension $\sigma$. Its contribution to the Euclidean instanton action is
\begin{equation}
    S_E = -\frac{1}{2\kappa^2}\int_{\mathrm{AdS}_5} \dd^5x\,\sqrt{g} \left(R + \frac{12}{\ell^2}\right) - \frac{1}{\kappa^2}\int_{\Sigma}\dd^4\xi\, \sqrt{h}\, K+\sigma \int_{\Sigma}\dd^4\xi\,\sqrt{h}\,,
\end{equation}
where $\Lambda = -6\ell^{-2}$ is the cosmological constant, and $\kappa^2=8\pi G_N^{(5)}$ the five-dimensional gravitational coupling. The bulk metric $g_{MN}$ with Ricci curvature $R_{MN}$ satisfies Einstein's equations with a negative cosmological constant
\begin{equation}
    R_{MN}-\frac{1}{2} R g_{MN}+\frac{6}{\ell^2} g_{MN}=0\,.
\end{equation}
In practice, $\Sigma$ is really just an ETW brane. The induced metric on $\Sigma$ is $h_{ij}$  and the extrinsic curvature $K_{ij}=\frac12 \mathcal{L}_n h_{ij}$, where $\mathcal{L}_n$  is the Lie derivative with respect to the outward unit normal, $n^M$.  Now, because there is nothing `outside' the bubble, we must impose boundary conditions at the bubble wall. Adopting Neumann boundary conditions, variation of the action with respect to the boundary metric at the bubble wall yields a one-sided version of the Israel junction conditions \cite{Israel:1966rt}, 
\begin{equation}\label{eq:junctiononesided}
    \frac{1}{\kappa^2} \left(K_{ij}-K h_{ij}\right)=- \sigma  h_{ij}\,.
\end{equation}
To find the instanton solution, we  make the following $O(5)$ symmetric ansatz
\begin{equation}\label{eq:onesidedmetric}
    \dd s^2 =  \dd r^2 + \chi(r)^2  \dd \Omega_4^2\,,
\end{equation}
where $\dd \Omega_4^2=\gamma_{ij} \dd \xi^i \dd \xi^j$ is metric on a unit $4$-sphere. The bulk Einstein equations yield a Euclidean AdS with $\chi(r)=\ell \sinh(r/\ell)$. The ETW brane is located at some $r=r_0$ and has unit normal $n^M \partial_M=\partial_r$ pointing in the direction of increasing volume. Note that the brane geometry is that of a $4$-sphere of radius $\chi(r_0)$, with induced metric $h_{ij}=\chi^2(r_0) \gamma_{ij}$. Its extrinsic curvature is readily computed to give
\begin{equation}\label{eq:extrinsiccurvature}
    K_{ij} = \frac{\chi'(r_0)}{\chi(r_0)}h_{ij}\,,
\end{equation}
where $'$ is the derivative with respect to the radial coordinate $r$. Combining this with Eq. \eqref{eq:junctiononesided} yields an explicit expression for $\sigma$. Notably, the brane tension is positive and superextremal.\footnote{This is slightly different from the dark bubble model, where the brane tension is subextremal. The superextremality here is tied to the fact that we can actually re-interpret our model as Randall-Sundrum like, which is discussed later towards the end of Section \ref{sec:renormalization}.} Namely, it is bounded from below, $\sigma>\sigma_\text{ext}$, where the extremal tension is that of a bubble of infinite size when compared to the AdS length.
\begin{equation}\label{eq:tensiononesided}
    \sigma = \frac{3}{\kappa^2\ell}\coth\left(\frac{r_0}{\ell}\right) \quad \xrightarrow{r_0\to\infty}\quad  \sigma_\text{ext} = \frac{3}{\kappa^2\ell}
\end{equation}
With these ingredients, as well as the solution to the bulk Einstein equations, $R=-20\ell^{-2}$, one can compute the Euclidean action of the bubble.
\begin{equation}\label{eq:bounceonesided}
    S_E = \frac{3\Omega_4 \ell^3}{2\kappa^2}\left[\frac{r_0}{\ell} - \sinh\left(\frac{r_0}{\ell}\right)\cosh\left(\frac{r_0}{\ell}\right)\right]
\end{equation}
Here we have pulled out an overall factor $\Omega_4\equiv \int \dd \Omega_4=\frac{8}{3} \pi^2$ that is the area of the unit 4-sphere. Note that this Euclidean action is always negative.

This Bubble of Something can be interpreted as the creation of an AdS$_5$ universe with a braneworld boundary, from nothing. Given that $S_E<0$, there are two proposals for the probability of creating such a universe, 
\begin{equation}
    P \propto e^{\pm \abs{S_E}}\,.
\end{equation}
The choice of sign corresponds to choosing suitable boundary conditions for the wavefunction of the universe. The upper sign corresponds to the choice of Hartle and Hawking \cite{Hartle:1983ai}, motivated by the Hartle-Hawking wavefunction and the no-boundary proposal. The lower sign is due to Linde \cite{Linde:1983mx} and Vilenkin \cite{Vilenkin:1984wp}, and is based on a tunnelling wavefunction. For a recent accessible discussion on these proposals in the context of slow roll inflation, see \cite{Maldacena:2024uhs}. As it happens, for the Hartle-Hawking  proposal, the dominant configurations are those with large Euclidean action in absolute value. This pushes $r_0$ to large values, or equivalently, the ETW brane to large radius and small curvature. Oppositely, the Linde-Vilenkin proposal will favour small values of the Euclidean action, pushing the brane towards a smaller radii. There remains to this day no consensus in the literature about which sign is the correct one to adopt, although in the context of the cosmological constant problem \cite{Weinberg:1988cp,Padilla:2015aaa,Burgess:2013ara}, we might prefer the Hartle-Hawking choice. However, we urge caution in drawing any premature conclusions. Part of the problem lies in the fact that the Euclidean action in Eq. \eqref{eq:bounceonesided} is unbounded from below. No matter whether one prefers Hartle-Hawking or Linde-Vilenkin, this means we are steered towards extreme conclusions: either $r_0/\ell\to \infty$ becomes absolutely favoured, or it is infinitely suppressed. Behind the scenes, the unboundedness is nothing but a manifestation of the famous conformal factor problem in Euclidean quantum gravity \cite{Gibbons:1978ac}. To see this more clearly, consider fixing $M_{\text{Pl},5}\ell$, which leaves the size of the bubble $r_0/\ell$ as the only relevant scale in the theory. Varying the size of the bubble then equivalently amounts to performing a conformal transformation. For these reasons, we do not find it reasonable to make claims about the preferred size or tension of the bubble. Nevertheless, it is important to remember that the tension will always be positive in our case, with the largest bubbles giving a small effective 4d cosmological constant on the ETW brane. 

The evolution of the ETW brane is obtained by Wick rotating the Euclidean solution back to Lorentzian signature. To preserve the asymptotic structure of the AdS vacuum, we need to do this on one of the coordinates parametrising the $S^4$. Thus $\dd \Omega_4^2 = \dd T^2 + \sin^2T\dd \Omega_3^2 \to -\dd \tau^2 + \cosh^2 \tau \dd \Omega_3^2$ (where $T = \pi/2 + i\tau$), which is precisely the global patch of de Sitter. It is glued to the Euclidean ball at $\tau=0$. The 5d metric becomes
\begin{equation}
    \dd s^2 = \dd r^2 + \chi^2(r) \left(-\dd \tau^2 + \cosh^2\tau \dd \Omega_3^2\right)\,.
\end{equation}
Defining the cosmological time coordinate $t = \chi(r_0) \tau = H_0^{-1}\tau$ yields in either coordinate patch
\begin{equation}\label{eq:desitterslice}
    \dd s^2 = \dd r^2 + \left(\frac{\sinh(r/\ell)}{\sinh(r_0/\ell)}\right)^2 \left[-\dd t^2 + H_0^{-2}\cosh^2(H_0t) \dd \Omega_3^2\right]\,.
\end{equation}
The line element inside the square bracket can be identified as the metric of $\mathrm{dS}_4$. The Hubble rate depends only on the radius $\chi(r_0)$ and for large ETW branes
\begin{equation}
    H_0 = \chi(r_0)^{-1}\sim \frac{2}{\ell} e^{-r/\ell} \ll \ell^{-1}\,.
\end{equation}
Given that this set-up is essentially half of a Randall-Sundrum like model, fluctuations around this solution will recover 4d gravity on the ETW brane at scales much larger than the AdS length scale in the bulk \cite{Randall:1999vf}. Therefore, large ETW branes could mimic the gravitational properties of our universe, described by 4d General Relativity to leading order, with a small effective cosmological constant.

\subsection{Two-sided AdS bubble nucleation}\label{sec:two-sidedAdS}
In the spirit of Brown and Dahlen \cite{Brown:2011gt}, we now consider an alternative perspective, that is the general decay between different AdS vacua,  $\mathrm{AdS}_5^+\to \mathrm{AdS}_5^-$, where $\ell_+$ denotes the AdS radius of the parent vacuum outside the bubble, and $\ell_-$ is that of the daughter vacuum in the interior. In the limit where $\ell_+\to 0$, we will now argue that this is equivalent to the one-sided setup described above, with an ETW brane as the domain wall. To demonstrate the equivalence, we will need to renormalize a divergent brane tension. Later, in Section \ref{sec:renormalization}, we will justify this through holographic renormalization \cite{Skenderis:2002wp, Papadimitriou:2016yit}.

The two AdS regions satisfy Einstein's equations with the corresponding cosmological constants. They can be described by different coordinate patches as 
\begin{equation}\label{eq:twosidedmetric}
    \dd s^2 = \dd r^2 + \chi_\pm^2(r)\dd \Omega_4^2
\end{equation}
where $\chi_\pm (r)=\ell_\pm \sinh(r/\ell_\pm)$. In the parent coordinates, the bubble is located at $r=r_+$, whilst in the daughter coordinates it is at $r=r_-$. The induced metric is $h_{ij}=\chi^2_\pm (r_\pm)\gamma_{ij}$. We take $r_\pm>0$ and in general $r_+\neq r_-$. The relation between the two radii follows from the fact that the induced metric on the brane must be well defined, giving the continuity condition $ \chi_+(r_+) = \chi_-(r_-)$, or in other words,
\begin{equation} \label{cont}
   \ell_+ \sinh\left(\frac{r_+}{\ell_+}\right)= \ell_-\sinh\left(\frac{r_-}{\ell_-}\right)\,.
\end{equation}
This should be supplemented by the Israel junction condition \cite{Israel:1966rt}, relating the jump in extrinsic curvature as we cross the brane of tension $\sigma$,
\begin{equation}
   - \frac{1}{\kappa^2} \left[K_{ij}-K h_{ij}\right]_-^+=- \sigma  h_{ij}\,.
\end{equation}
Here we denote $[Q]^+_-=Q_+-Q_-$ and $K_{ij}^\pm=\frac{\chi_\pm'(r_\pm)}{\chi_\pm(r_\pm)}h_{ij}$ is the extrinsic curvature of the brane in the corresponding AdS$_5^\pm$ bulk spacetime. Plugging in the explicit form for the jump in extrinsic curvature for the instanton solution, we obtain 
\begin{equation} \label{tension}
 \sigma=\frac{3}{\kappa^2} \left[ \frac{1}{\ell_-}\coth\left(\frac{r_-}{\ell_-}\right) - \frac{1}{\ell_+}\coth\left(\frac{r_+}{\ell_+}\right)\right]\,.
\end{equation}
For up-tunnelling $\ell_+<\ell_-$, the continuity condition \eqref{cont} implies that $r_-/\ell_-<r_+/\ell_+$ and so $\sigma<0$. This negative tension is unphysical and the reason that up-tunnelling between AdS vacua in General Relativity is ruled out (see \cite{Liu:2024aos} for a loophole in Gauss-Bonnet gravity). Our interest here lies in the limiting case $\ell_+\to 0$ where the tension diverges towards minus infinity $\sigma \to -\infty$. More precisely, if $\chi_b\equiv \chi_+(r_+) = \chi_-(r_-)$ is the radius of the $4$-sphere at the brane, we can rewrite Eq. \eqref{tension} as
\begin{equation}
 \sigma=\frac{3}{\kappa^2} \left[ \frac{1}{\ell_-}\coth\left(\frac{r_-}{\ell_-}\right) - \frac{1}{\ell_+}\sqrt{1+\frac{\ell_+^2}{\chi_b^2}}\right]\,.
\end{equation}
In the limit where $\ell_+\to 0$, holding $\chi_b$ fixed, we see that we can identify, 
\begin{equation}\label{eq:baretensionsigma}
    \sigma = \sigma_\text{bare} = \sigma_\text{ren}+\sigma_\text{div}
 \end{equation}   
where 
\begin{equation} \label{eq:sigmadiv}
    \sigma_\text{div} =-\frac{3}{\kappa^2 \ell_+}
\end{equation}
is the divergent piece and 
\begin{equation} \label{sigmaren}
 \sigma_\text{ren} = \frac{3}{\kappa^2\ell_- } \coth\left(\frac{r_-}{\ell_-}\right)\,.
\end{equation}
Now, as $\ell_+\to 0$, it is natural to associate the daughter vacuum with the bulk geometry in the one-sided case discussed in the previous section, identifying $r_-$ with $r_0$, $\chi_-(r)$ with $\chi(r)$ and $\ell_-$ with $\ell$ and so on. If we do this, we immediately see that $\sigma_\text{ren}$ is just the positive tension of the ETW brane that we found in Eq. \eqref{eq:tensiononesided}. 

With some foresight we are alluding to $\sigma_\text{bare}$ and $\sigma_\text{ren}$ as being respectively some ``bare'' and ``renormalized'' tensions. The same divergent behaviour was also discussed in \cite{Friedrich:2023tid}.\footnote{In \cite{Friedrich:2023tid} the authors discuss Bubbles of Something with a $d=4$ dimensional bulk instead of $d=5$. They find a divergent behaviour $\sigma_\text{div}= -\frac{2}{\kappa^2\ell_+}$. We have confirmed that for a general $d$ the numerator is $d-2$ and that our results apply to their setup as well.} Here the authors argue that a Bubble of Something realised by up-tunnelling from AdS with $\ell_+\to 0$ requires one to take $\sigma_\text{bare}\to -\infty$ simultaneously, such that $\sigma_\text{ren}$ is kept finite. In the next section we show precisely how to realise this renormalization. Our interpretation is as follows: the infinite negative tension is really an artefact of the action not being renormalized near the conformal boundary. Divergences of such origin can be mended through holographic renormalization, which introduces counterterms to the effective action on the boundary. We will show that the appropriate counterterm one shall add kills the divergence in Eq. \eqref{eq:sigmadiv}. This conclusion will also hold true in any $d$ dimensions.

We now turn our attention to the decay rate for the transition between vacua in the two-sided case. In the semi-classical theory of vacuum decay \cite{Coleman:1977py, Callan:1977pt, Coleman:1980aw}, this is given by 
\begin{equation}
    \frac{\Gamma}{\text{Vol}} \propto e^{-B}\,,
\end{equation}
where the exponent 
\begin{equation}
    B=S_E[\text{instanton}]-S_E[\text{parent}]
\end{equation}
is the difference between the Euclidean actions of the instanton describing the transition and the parent vacuum. In computing $B$, we need to be slightly careful as both the parent and instanton Euclidean actions involve radial integrals that diverge as they integrate up to the conformal boundary at infinite radius. Take for instance the parent action. This can be dissected into a Euclidean action for the parent interior and a parent exterior, separated by the surface at $r=r_+$,
\begin{equation}\label{eq:parenttwosided}
    S_E[\text{parent}] = S_E^\text{in}[\text{parent}]+S_E^\text{out} [\text{parent}]\,,
\end{equation}
where 
\begin{eqnarray}
 S_E^\text{in}[\text{parent}]&=&-\frac{1}{2\kappa^2}\int_{r\leq r_+} \dd^5x\, \sqrt{g} \left(R+\frac{12}{\ell_+^2}\right)- \frac{1}{\kappa^2}\int_{r=r_+}  \dd^4 \xi\, \sqrt{h}\, K_+ \\
S_E^\text{out}[\text{parent}]&=&-\frac{1}{2\kappa^2}\int_{r\geq r_+} \dd^5x\, \sqrt{g} \left(R+\frac{12}{\ell_+^2}\right)+\frac{1}{\kappa^2} \int_{r=r_+}  \dd^4 \xi\, \sqrt{h}\, K_+ +\cdots \label{Soutparent}
\end{eqnarray}
and the ellipses denote terms defined on the conformal boundary. The exterior contribution, $S^\text{out}_E[\text{parent}]$, is at risk of being divergent. As we will see in the next section, one can regularize these integrals as we approach the conformal boundary and add boundary counterterms that absorb these divergences. In principle, we could be explicit about these terms included in the ellipses. However, it is not necessary to do this explicitly at this stage, as the instanton action contains the exact same divergences. In particular, we find that
\begin{equation}\label{eq:instantontwosided}
    S_E[\text{instanton}] = S_E^\text{in}[\text{daughter}]+S_E[\text{brane}]+S_E^\text{out} [\text{parent}]\,,
\end{equation}
where $S_E^\text{out} [\text{parent}]$ was given in Eq. \eqref{Soutparent} and 
\begin{eqnarray}
 S_E^\text{in}[\text{daughter}]&=&-\frac{1}{2\kappa^2}\int_{r\leq r_-} \dd^5x\, \sqrt{g} \left(R+\frac{12}{\ell_-^2}\right)-\frac{1}{\kappa^2} \int_{r=r_-}  \dd^4 \xi\, \sqrt{h}\, K_-\,, \\
S_E[\text{brane}]&=& \sigma \int_{r=r_-} \dd^4 \xi\, \sqrt{h}\,.  \label{SEbrane}
\end{eqnarray}
When we compute the tunnelling exponent $B$, we see that the divergences coming from the conformal boundary exactly cancel, giving 
\begin{equation}
    B=S_E^\text{in}[\text{daughter}]-S_E^\text{in}[\text{parent}]+S_E[\text{brane}]\,.
\end{equation}
After explicitly performing the integrals, we find that 
\begin{equation}\label{eq:bouncetwosided}
    B =\frac{3\Omega_4 }{2\kappa^2}\left[ \ell_-^3\left(\frac{r_-}{\ell_-} - \sinh\left(\frac{r_-}{\ell_-}\right)\cosh\left(\frac{r_-}{\ell_-}\right)\right)-\ell_+^3\left(\sinh^{-1}\left(\frac{\chi_b}{\ell_+}\right)-\frac{\chi_b}{\ell_+}\sqrt{1+\frac{\chi_b^2}{\ell_+^2}}\right)\right]\,.
\end{equation}
Remarkably, as we take the limit $\ell_+ \to 0$ holding $\chi_b$ fixed, we obtain the Euclidean action, Eq. \eqref{eq:bounceonesided}, after identifying the daughter vacuum with the bulk geometry in the one-sided case. Employing the Hamiltonian formalism instead of CdL to analyse vacuum transitions via bubble nucleation further supports this result \cite{Cespedes:2023jdk}. In this limit, the tunnelling rate obtained from semi-classical methods scales like the Hartle-Hawking formula for the probability of creating the AdS$_5$ vacuum with an ETW brane, from nothing.

\section{Holographic Renormalization}\label{sec:renormalization}
In this section we will show that the two interpretations of the bubble of AdS from nothing, described in Section \ref{sec:bubbleofsomething}, can be made fully consistent with one another. Our resolution relies on the holographic duality between the gravitational theory in the AdS bulk and the CFT living on the conformal boundary \cite{Maldacena:1997re}. For asymptotically AdS spacetimes, the AdS/CFT correspondence \cite{Witten:1998qj} states that the partition function for bulk fields $\Phi$ with boundary value $\phi_0$ is identified with  the generating functional of the boundary CFT, with $\phi_0$ serving as sources for dual operators, $\mathcal{O}$,
\begin{equation}
   \mathcal{Z}_\text{SUGRA}[\phi_0]=\int_{\Phi 
   \sim \phi_0} \mathcal{D} \Phi\, e^{-S[\Phi]}=\left\langle e^{-\int \phi_0\mathcal{O}}\right\rangle_\text{CFT} \,.
\end{equation}
On the CFT side, it is well-appreciated that we should employ renormalization techniques in order to get rid of any UV divergences and make further sense of the theory. Holography tells us that they are to be mapped to IR divergences in the bulk, i.e. as one approaches the conformal boundary. In fact, we already saw this in the previous section; the on-shell Euclidean action admits divergences associated to the infinite volume of AdS. The procedure to render them finite is dubbed holographic renormalization. See for instance \cite{Skenderis:2002wp, Papadimitriou:2016yit} for pedagogical introductions to the topic. The strategy essentially amounts to expanding the bulk action near the conformal boundary whereafter one can read off the counterterms required to render the Euclidean action finite. We will make use of some well-known results that can be found in \cite{Kraus:1999di, deHaro:2000vlm}, but let us nevertheless provide a minimal introduction to provide some context for our main results. For the remainder of this paper, we work in Lorentzian signature.

In the coordinates we have used so far, the conformal boundary lies beyond the bubble at $r\to \infty$. However, these are not the most useful coordinates for understanding the theory in this radial limit. Instead, it is standard to rewrite asymptotically AdS spacetimes in terms of Fefferman-Graham (FG) coordinates
\begin{equation} \label{FG}
    \dd s^2 = \frac{\ell^2}{4\varrho^2}\dd\varrho^2 + \frac{\ell^2}{\varrho}\hat{g}_{ij}(\varrho,\xi)\dd \xi^i\dd \xi^j
\end{equation}
where $ \varrho = e^{-2r/\ell}$ and 
\begin{equation}
    \hat{g}_{ij}(\varrho,\xi) = \hat{g}_{ij}^{(0)}(\xi) + \varrho\hat{g}_{ij}^{(2)}(\xi) + \varrho^2\left(\hat{g}_{ij}^{(4)}(\xi) + \log(\varrho) \hat{h}_{ij}^{(4)}(\xi)\right) + \mathcal{O}(\varrho^3)\,.
\end{equation}
In these new coordinates, the conformal boundary is located at $\varrho\to 0$ and has metric $\hat{g}_{ij}^{(0)}$. The following term $\hat{g}_{ij}^{(2)}$ and the trace of $\hat{g}_{ij}^{(4)}$ is completely determined by $\hat{g}^{(0)}_{ij}$ and its associated covariant derivative \cite{deHaro:2000vlm}. However the transverse-traceless part of $\hat{g}_{ij}^{(4)}$ is in general not fixed and depends on the choice of boundary conditions \cite{Papadimitriou:2004ap}. This fact will become important when we later discuss the linearised perturbations on the bubble. Finally, $\hat{h}_{ij}^{(4)}$, which is accompanied by the logarithm, only appears when the boundary dimension is even, and is related to the holographic Weyl anomaly \cite{Henningson:1998gx, Henningson:1998ey}. 

The on-shell action for  gravity diverges at the conformal boundary and must be renormalized. To see this, we  introduce a radial cutoff at some $\varrho=\varepsilon\ll 1$ in FG coordinates, so that the gravitational action reads
\begin{equation}
    S_{\text{reg}, \varepsilon}=\frac{1}{2\kappa^2}\int_{\varrho>\varepsilon} \dd^5x\, \sqrt{-g} \left(R + \frac{12}{\ell^2}\right) + \frac{1}{\kappa^2}\int_{\varrho=\varepsilon}\dd^4\xi\, \sqrt{-h}\, K\,.
\end{equation}
This yields a regularised action containing a finite number of terms that diverge in the limit $\varepsilon\to 0$, of the form
\begin{equation}\label{eq:Sregepsilon}
    S_\text{reg,$\varepsilon$} = \int_{\varrho=\varepsilon} \dd^4\xi\, \sqrt{-\hat{g}^{(0)}} \left(\varepsilon^{-2}\hat{\mathcal{L}}^{(0)} + \varepsilon^{-1}\hat{\mathcal{L}}^{(2)} - \log(\varepsilon) \hat{\mathcal{L}}^{(4)} + \mathcal{O}(\varepsilon^0)\right)\,.
\end{equation}
The $\hat{\mathcal{L}}^{(2n)}$ are covariant objects of $\hat{g}^{(0)}_{ij}$ to be determined. These are the ones that are to be removed by adding a counterterm action on the boundary,
\begin{equation}
    S_\text{ct,$\varepsilon$} \equiv -\mathrm{div}\left(S_\text{reg,$\varepsilon$}\right) = -\int_{\varrho=\varepsilon} \dd^4\xi \, \sqrt{-\hat{g}^{(0)}}\left(\varepsilon^{-2}\hat{\mathcal{L}}^{(0)} + \varepsilon^{-1}\hat{\mathcal{L}}^{(2)}-\log(\varepsilon)\hat{\mathcal{L}}^{(4)}\right)\,.
\end{equation}
In the end we get a renormalized action that  is well-behaved in the limit $\varepsilon\to 0$, where the cutoff brane is effectively removed.
\begin{equation}
    S_\text{ren} \equiv \lim_{\varepsilon\to 0}S_\text{ren,$\varepsilon$}\,,\qquad S_\text{ren,$\varepsilon$}\equiv S_\text{reg,$\varepsilon$} + S_\text{ct,$\varepsilon$}
\end{equation}
It is from this renormalized action that we are able to compute correlation functions of boundary operators. For instance, the one-point function of the renormalized energy momentum tensor of the boundary CFT is obtained from the variation with respect to the boundary metric $\hat{g}^{(0)}_{ij}$. However our final goal is not to describe the theory on the conformal boundary, but rather at some hypersurface of fixed $\varrho$ where the bubble is located. Of course, the bubble separates two spacetimes, labelled with $\pm$.  The exterior $+$ spacetime has $\varrho<\varrho_+$ with the bubble wall at $\varrho=\varrho_+$. The interior $-$ spacetime has $\varrho>\varrho_-$ with the bubble wall at $\varrho=\varrho_-$. It is useful to refer to $\varrho_-$ as the size of the bubble, taking the perspective of the interior spacetime. 

The induced metric on the bubble wall in FG coordinates is simply
\begin{equation}\label{eq:braneinducedmetricFG}
    h_{ij}(\xi) \equiv \frac{\ell_-^2}{\varrho_-}\hat{g}_{ij}^{(-)}(\varrho_-,\xi) = \frac{\ell_+^2}{\varrho_+}\hat{g}_{ij}^{(+)}(\varrho_+,\xi)\,.
\end{equation}
In principle, this continuity condition gives an expression for $\varrho_+$  as a function of $\varrho_-$, though the exact form will not be important to us. The extrinsic curvature of the bubble in the $\pm$ spacetime is given by 
\begin{equation}\label{eq:Kijpm}
    K^\pm_{ij}=-\frac{\varrho}{\ell_\pm} \pdv{}{\varrho}\left[\frac{\ell_\pm^2}{\varrho}\hat{g}_{ij}^{(\pm)}(\varrho,\xi)\right]\bigg\vert_{\varrho=\varrho_\pm}\,.
\end{equation}
Now we are ready to derive the renormalized action for the two-sided AdS bubble described in the previous section. It is just the Lorentzian version of the instanton action and is given by
\begin{equation}\label{eq:bubbleactiondecomposition}
    S_\text{bubble} = S_\text{in}[-] + S_\text{bare}[\text{brane}] + S_\text{out}[+]\,.
\end{equation}
where 
\begin{eqnarray}
 S_\text{in}[\pm]&=&\frac{1}{2\kappa^2}\int_{\varrho> \varrho_\pm} \dd^5x\, \sqrt{-g} \left(R+\frac{12}{\ell_\pm^2}\right) + \frac{1}{\kappa^2}\int_{\varrho=\varrho_\pm}  \dd^4 \xi\, \sqrt{-h}\, K^\pm \\
 S_\text{out}[\pm]&=& \frac{1}{2\kappa^2}\int_{\varrho< \varrho_\pm} \dd^5x\, \sqrt{-g} \left(R+\frac{12}{\ell_\pm^2}\right) - \frac{1}{\kappa^2}\int_{\varrho=\varrho_\pm}  \dd^4 \xi\, \sqrt{-h}\, K^\pm +\cdots
\end{eqnarray}
and 
\begin{equation} \label{brane}
    S_\text{bare}[\text{brane}]= \int \dd^4 \xi \, \sqrt{-h} \left(-\sigma_\text{bare}+\mathcal{L}_m\right)
\end{equation}
A few comments are in order here. As in the previous section, the ellipses in the exterior action, $S_\text{out}[\pm]$, include the terms defined on the (regulated) conformal boundary, including counterterms that absorb the divergences. This is the renormalization procedure we have just described, and so we note that  
\begin{equation} \label{split}
S_\text{out}[\pm]= S_\text{ren}[\pm]-S_\text{in}[\pm]\,,
\end{equation}
where $S_\text{ren}[\pm]$ the renormalized action for the full $\pm$ spacetime. Eq. \eqref{split} is simply the statement that one can split the full renormalized action into an interior and a renormalized exterior. The interior action does not require any renormalization since it is already finite -- at least for finite values of $\ell_\pm$. This is because it is just the regulated action with a cutoff at $\varrho=\varrho_\pm$, or in other words,  $S_\text{in}[\pm]=S_\text{reg,$\varrho_\pm$}[\pm]$. The bare action for the brane, Eq. \eqref{brane}, contains the bare tension, $\sigma_\text{bare}$, and the bare Lagrangian for any localized matter excitations  on the brane, $\mathcal{L}_m$.   

It is important to highlight that $\varrho_\pm$, as cutoffs, receive special treatment compared to the usual holographic renormalization recipe. Indeed, we would usually renormalize by placing a cutoff at some $\varepsilon$, add counterterms, and subsequently send $\varepsilon\to 0$. But $\varrho_\pm$ are not some arbitrary quantities to later be discarded. In principle, they are finite physical scales that influence the effective physics on the domain wall. There are, however, subtleties that emerge as we take $\ell_+ \to 0$, holding $\ell_-$ and $\varrho_-$ fixed. In this limit, $\varrho_+ \to 0$ \footnote{
This is readily shown using our reparameterization to FG coordinates as well as Eq. \eqref{cont}. The leading order behaviour of $\varrho_+$ in terms of $\ell_+$ and $H_0$ is 
\begin{equation}
    \varrho_+ = e^{-2r_+/\ell_+} = \left(\frac{\ell_+H_0}{2}\right)^2\left(1+\mathcal{O}\left(\ell_+^2H_0^2\right)\right) \xrightarrow[H_0 \text{ fixed}]{\ell_+\to 0} 0\,.
\end{equation}
} and so it indeed becomes more appropriate to work with a renormalized action in the $+$ spacetime, 
\begin{equation}
    S_\text{ren,$\varrho_+$}[+]=S_\text{reg,$\varrho_+$}[+]+S_\text{ct,$\varrho_+$}[+]\,.
\end{equation}
In contrast, there is no need to introduce a similar renormalized action in the $-$ spacetime $S_\text{ren,$\varrho_-$}[-]$, since we never take $\varrho_- \to 0$ and $S_\text{reg,$\varrho_-$}[-]$ is always finite. This is unlike the holographic treatment of the dark bubble in \cite{Banerjee:2023uto}, where the renormalization scheme was run from both sides of the bubble simultaneously.

Bringing everything together, we note that 
\begin{equation} \label{split2}
\begin{split}
    S_\text{out}[+] &= S_\text{ren}[+]-S_\text{reg,$\varrho_+$}[+]\\
    &=S_\text{ren}[+]-S_\text{ren,$\varrho_+$}[+]+S_\text{ct,$\varrho_+$}[+]
\end{split}
\end{equation}
so that  Eq. \eqref{eq:bubbleactiondecomposition} becomes 
\begin{equation}\label{eq:Sbubbleren}
    S_\text{bubble} = S_\text{in}[-] +S_\text{ren}[\text{brane}]+ \Delta S_\text{ren}[+]\,, 
\end{equation}
where 
\begin{equation}
    \Delta S_\text{ren}[+] \equiv S_\text{ren}[+]-S_\text{ren,$\varrho_+$}[+]
\end{equation}
and we have introduced the following renormalized action for the brane
\begin{equation}
    S_\text{ren}[\text{brane}] \equiv S_\text{bare}[\text{brane}]+S_\text{ct,$\varrho_+$}[+]\,.
\end{equation}
Let us digest the terms appearing in this rewritten version of the action. First off, $S_\text{in}[-]$ remains unchanged, and is the bulk contribution from inside the bubble. From a holographic perspective, it is the gravitational dual of a UV-cutoff CFT living on the bubble. Next, $S_\text{ren} [\text{brane}]$ gives the induced gravitational action on the brane. When $\ell_+\neq 0$ the counterterms appear as higher-curvature corrections to the action. The fact that counterterms diverge as $\varrho_+\to 0$ signals an increasing localization of gravity on the brane \cite{Bueno:2022log}. Finally, $\Delta S_\text{ren}[+]$ is an entirely finite action describing bulk contributions from outside the bubble. We expect it to vanish completely in the limit $\ell_+\to 0$, since  $\varrho_+ \to 0$ and so $S_\text{ren,$\varrho_+$} \to S_\text{ren}$ by definition. 

\subsection{Algorithm for computing counterterms}
To determine the counterterms starting from the bulk action, one could explicitly integrate out $\varrho$ and identify the divergent terms that are to be minimally subtracted. This turns out to be a laborious task -- especially because the counterterms in this way yield an action on the conformal boundary, whereas we are interested in an action on the bubble. Luckily there exist straightforward algorithms to compute these counterterms iteratively in orders of $\ell$. Here we closely follow the method developed in \cite{Kraus:1999di} which has also been applied in a similar fashion in \cite{Bueno:2022log}. We focus on the case where the boundary dimension is $d=4$.

The intuitive picture to have in mind is the following: counterterms associated to the outside bulk effectively imprint a (generally) higher curvature gravity theory on the bubble. On the other hand, the bulk theory from inside the bubble is holographically dual to a cutoff CFT. In this way, the bubble observer who does not immediately realise the physics in the bulk, experiences a gravitationally coupled holographic QFT. 

Now we focus on the outside contribution. It is useful to think in terms of the associated boundary energy momentum tensor
\begin{equation}\label{eq:Piboundaryenergymomentum}
    \Pi_{ij} = -\frac{2}{\sqrt{-h}}\frac{\delta}{\delta h^{ij}} \int \dd^4 \xi\, \sqrt{-h}\, \mathcal{L}
\end{equation}
with $\mathcal{L}$ some Lagrangian obtained from regularising $S[+]$ along the lines of Eq. \eqref{eq:Sregepsilon}. The idea is then to extract counterterms that cancel the divergences, $\widetilde{\Pi}_{ij} = -\mathrm{div}(\Pi_{ij})$, which we argued from Eq. \eqref{eq:Sbubbleren} is what gets imprinted on the bubble. They can be written as a curvature expansion
\begin{equation}\label{eq:PiLexpansion}
    \Pi_{ij} = \Pi_{ij}^{(0)} + \Pi_{ij}^{(1)} + \cdots\,,\qquad \mathcal{L}=\mathcal{L}^{(0)}+\mathcal{L}^{(1)}+\cdots\,,
\end{equation}
where $\mathcal{L}^{(n)}\propto \ell_+^{2n-1}R^n[h]$. Moreover, one can show that $\Pi^{(n)}=(4-2n)\mathcal{L}^{(n)}$ by looking at the properties of Eq. \eqref{eq:Piboundaryenergymomentum} under local Weyl transformations \cite{Kraus:1999di}. This holds up to total derivatives and is valid as long as $\mathcal{L}$ is local. On the other hand, it clearly breaks down for $n=2$ due to the holographic Weyl anomaly, which we recall, also involves a non-local logarithmic piece. In order to recover the correct result, it was shown that it is sufficient to replace $(4-2n)$ with $(\log\varrho_+)^{-1}$ whenever we would have to insert $n=2$. Next, consider the Gauss-Codazzi equations, which correspond to solving the bulk Einstein equations of $S[+]$ at the bubble. Here we use the scalar constraint
\begin{equation}
    G_{MN}[g]n^M n^N = \frac{1}{2}\left(\frac{1}{3}\kappa^4\Pi^2 - \kappa^4\Pi_{ij}\Pi^{ij}-R[h]\right)
\end{equation}
where $\Pi=h^{ij} \Pi_{ij}$. The left-hand side gives on-shell the usual $6\ell^{-2}_+$ from the bulk vacuum equation. Using the expansion of Eq. \eqref{eq:PiLexpansion} it is now possible to solve for $\Pi$ and hence extract $\mathcal{L}$ order by order in $\ell_+$.
\begin{equation}
\begin{split}
    \mathcal{L}^{(0)} &= \frac{3}{\kappa^2\ell_+}\\
    \mathcal{L}^{(1)} &= \frac{\ell_+}{4\kappa^2}R[h]\\
    \mathcal{L}^{(2)} &= \frac{\ell_+^3\log\varrho_+}{8\kappa^2}\left(R_{ij}R^{ij}-\frac{1}{3}R^2\right)
\end{split}
\end{equation}
The counterterm Lagrangian is precisely the sum of these three leading terms but with an opposite overall sign. The respective action gives
\begin{equation}
\begin{split}
    S_\text{ct,$\varrho_+$}[+] &= -\int\dd^4\xi\, \sqrt{-h} \left(\mathcal{L}^{(0)}+\mathcal{L}^{(1)}+\mathcal{L}^{(2)}\right)\,,\\
    &= -\frac{\ell_+}{\kappa^2}\int \dd^4\xi\, \sqrt{-h}\left[\frac{3}{\ell_+^2}+\frac{1}{4}R+\frac{\ell_+^2\log\varrho_+}{8}\left(R_{ij}R^{ij}-\frac{1}{3}R^2\right)\right]\,.
\end{split}
\end{equation}
Recall then that the renormalized brane action is given by $S_\text{ren}[\text{brane}]=S_\text{bare}[\text{brane}]+S_\text{ct,$\varrho_+$}[+]$ with the bare action involving the bare tension in Eq. \eqref{eq:baretensionsigma}. If $\ell_+\neq 0$, a localized observer on the bubble would feel the effects of the exterior spacetime through these gravitational corrections. The sign of the Einstein-Hilbert coupling is negative, reflecting the fact that gravitational fluctuations localize near the conformal boundary, rather than at the brane. However, let us explore what happens to $S_\text{ren}[\text{brane}]$ when taking the limit $\ell_+\to 0$. What we find is that all curvature counterterms vanish and the part that remains surgically eliminates the divergence contained in the tension.
\begin{equation}
\begin{split}
    \lim_{\ell_+\to 0}S_\text{ren}[\text{brane}] &= \lim_{\ell_+\to 0} \int \dd^4\xi \, \sqrt{-h} \left[ -\left( \sigma_\text{ren} + \sigma_\text{div} \right)+\mathcal{L}_m - \frac{3}{\kappa^2\ell_+} + \mathcal{O}(\ell_+) \right]\\
    &=\int \dd^4\xi \, \sqrt{-h}\, ( -\sigma_\text{ren}+\mathcal{L}_m)
\end{split}
\end{equation}
Here we have used Eqs. \eqref{eq:baretensionsigma} and \eqref{eq:sigmadiv}. Thus, the divergent tension that is apparent when one considers the decay $\mathrm{AdS}^+_5\to\mathrm{AdS}^-_5$ with $\ell_+\to 0$ is just an artefact of the action not being renormalized. After performing holographic renormalization, the effective tension is both finite and positive, $\sigma_\text{ren}>0$, and an observer on the bubble sees a dS spacetime. The effective cosmological constant is small for large bubbles.

In total, the renormalized bubble action in the limit $\ell_+\to 0$ effectively reduces to
\begin{equation}\label{eq:Sbubblerenlimit}
    \lim_{\ell_+\to 0}S_\text{bubble} = S_\text{in}[-] + \int \dd^4\xi\, \sqrt{-h}\, (-\sigma_\text{ren}+\mathcal{L}_m)\,,
\end{equation}
which we interpret as the total absence of spacetime outside the bubble. In other words, the bubble effectively becomes an ETW brane in this limit. It further motivates the proposal by Brown and Dahlen that `nothing' can be thought of as AdS with $\Lambda\to -\infty$. However, what we also demonstrate is that ETW branes via Bubbles of Something can be a way to obtain a dS cosmology, possibly even from string theory. Holographic renormalization is in fact the missing link if we want to think of this as the up-tunnelling from AdS with $\Lambda\to -\infty$.

\begin{figure}[ht]
  \centering  
  \includegraphics[width=0.9\textwidth]{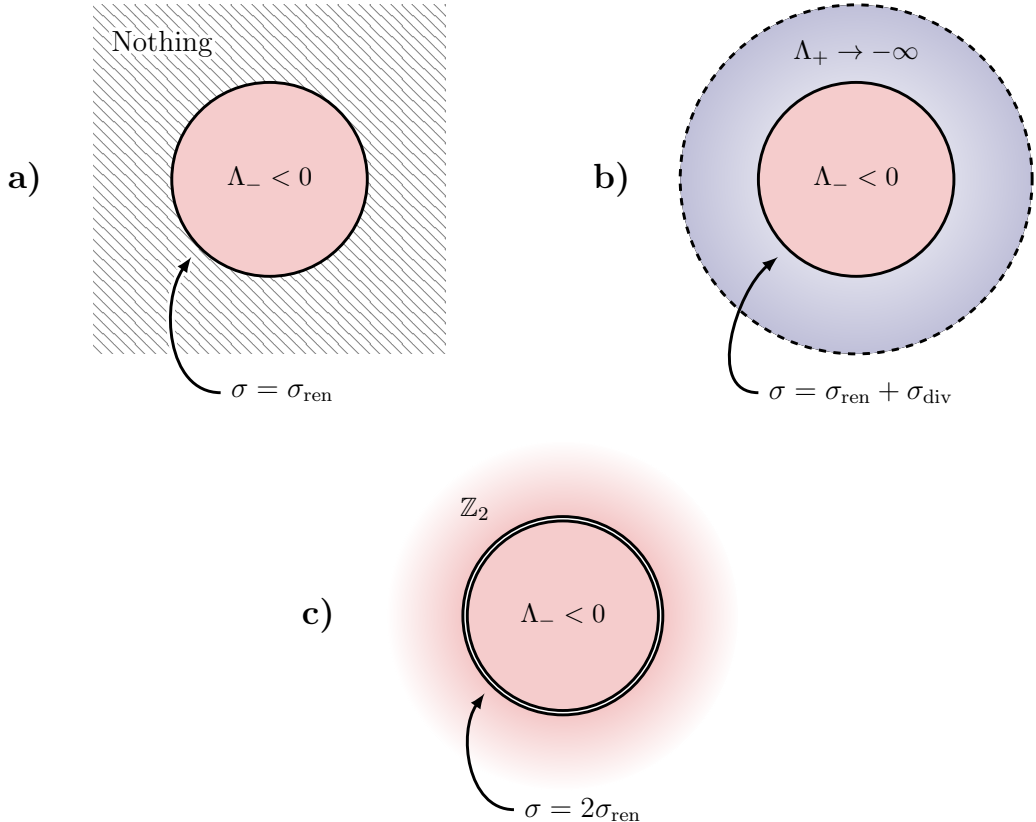}
  \caption{Illustrating the three perspectives of the bubble of something: \textbf{a)} the common one-sided picture of a finite volume of AdS where spacetime ends past the bubble wall. \textbf{b)} replace `nothing' outside with AdS with vacuum energy $\Lambda_+\to -\infty$. This two-sided picture leads to an infinite negative tension domain wall. But as we argue, this infinity can be isolated and cancelled by holographic renormalization. \textbf{c)} gluing two bubbles together and thereby imposing $\mathbb{Z}_2$ symmetry leads to an Randall-Sundrum like braneworld carrying twice the finite tension.}
    \label{fig:threebubbles}
\end{figure}

Finally, there is a third way of interpreting Eq. \eqref{eq:Sbubblerenlimit}, which we already alluded to in Section \ref{sec:bubbleofsomething}. Taking two copies of the bubble and gluing them along their boundaries gives the same action as a Randall-Sundrum like model, where the brane tension is twice that of the ETW brane, $\sigma_\text{RS}=2\sigma_\text{ren}$. In Euclidean signature this has a nice topological picture: the two inner bulks are each topologically equivalent to the 5-dimensional disc. Gluing them along the bubble that is the $S^4$ boundary can be seen in one dimension higher as gluing the two hemispheres of $S^5$. Therefore, we could consider our bubble as a Randall-Sundrum like model by performing a pushout to $S^5$.  The three different perspectives are summarized in Figure \ref{fig:threebubbles}.

\subsection{A closer look at perturbations}\label{sec:perturbations}
The above analysis applies quite generally and explicitly demonstrates the equivalence between a one-sided Randall-Sundrum like model and a two-sided setup. In the first case, the braneworld is an ETW brane. In the second picture, the braneworld is a bubble separating interior and exterior spacetimes endowed with respectively a finite and infinitely negative cosmological constant. The key to matching the two descriptions is holographic renormalization. However, this equivalence immediately prompts a puzzle. Fluctuations around the vacuum solution in the one-sided case closely follow those of a standard Randall-Sundrum setup (for a review of fluctuations in Randall-Sundrum gravity, see, for example \cite{Padilla:2002tg}). 
In particular, since the bulk volume is finite, there exists a normalizable graviton zero mode, implying that an observer on the ETW brane will recover 4d gravity at low energies. In contrast, for the two-sided case, the exterior spacetime includes the conformal boundary of AdS, making the bulk spacetime volume infinite. Here there is no normalizable graviton zero mode. One would therefore argue that an observer living on the bubble does not recover 4d gravity at low energies. There seems to be a contradiction?

At this point it is useful to note that the absence of a normalizable graviton zero mode is also a problem for the dark bubble. There, the problem is solved by introducing sources on the conformal boundary in the exterior spacetime \cite{Banerjee:2020wov}. To understand this in a broader sense, let us digress from bubbles and branes for the moment and consider some general asymptotically AdS spacetime, $\mathcal{M}$,  described by a renormalized action $S_\text{ren}[\mathcal{M}]$. The vacuum solution is an AdS spacetime with an AdS length scale $\ell$. Now consider bulk perturbations,  $\delta g_{MN}$, which admit an FG expansion near the conformal boundary. We identify the transverse-traceless parts of  $\delta \hat{g}_{ij}^{(0)}$ and $\delta \hat{g}_{ij}^{(4)}$ as the non-normalizable and normalizable zero modes, which come with factors $\varrho^{-1}$ and $\varrho$ respectively. The normalizable mode vanishes on the conformal boundary as $\varrho\to 0$, while the non-normalizable mode does not. Of course, the non-normalizable mode is also not uniquely defined, since superposing it with a normalizable mode makes a new non-normalizable mode. From a holographic perspective, the non-normalizable mode is supported by sources in the boundary CFT. More precisely, there is a non-vanishing one-point function
\begin{equation}\label{eq:expvalTij}
    \expval{\mathcal{T}_{ij}}=-\frac{2}{\sqrt{-\hat{g}^{(0)}}}\frac{\delta S_\text{ren}}{\delta \hat{g}^{(0)ij}}\,.
\end{equation}
For pure gravity, this is well known and has been computed in \cite{deHaro:2000vlm} to be
\begin{equation} \label{onepoint}
    \expval{\mathcal{T}_{ij}} = \frac{\ell^3}{\kappa^2}\left[ 2\hat{g}^{(4)}_{ij} - \frac{1}{4}\hat{g}^{(0)}_{ij} \left[(\tr \hat{g}^{(2)})^2-\tr  (\hat{g}^{(2)})^2\right]  - \left(\hat{g}^{(2)}\right)^2_{ij} + \frac{1}{2}\hat{g}_{ij}^{(2)}\tr \hat{g}^{(2)}
    \right]\,.
\end{equation}
Note that indices are raised and lowered with respect to the boundary metric $\hat g^{(0)}_{ij}$, such that $\tr \hat{g}^{(2)}=\hat g^{(0)ij}\hat{g}_{ij}^{(2)}$ and $\left(\hat{g}^{(2)}\right)^2_{ij}=\hat{g}_{ik}^{(2)}\hat g^{(0)kl}\hat{g}_{lj}^{(2)}$, and so on. The bulk equations of motion can be used to show that $\hat{g}^{(2)}_{ij}$ is given by the Schouten tensor for $\hat{g}^{(0)}_{ij}$ \cite{deHaro:2000vlm},
\begin{equation}\label{eq:schouten}
    \hat g^{(2)}_{ij}=-\frac{1}{2} \left(R_{ij}[\hat g^{(0)}]-\frac16 R [\hat g^{(0)}] \hat g^{(0)}_{ij} \right)\,.
\end{equation}
Now we return our attention to branes and bubbles. In doing so, we need to clarify some of the language used in the previous paragraph. The identification of the transverse-traceless parts of $\delta \hat{g}_{ij}^{(0)}$ and $\delta \hat{g}_{ij}^{(4)}$ as the non-normalizable and normalizable zero modes is rather standard within the holographic community. However, this terminology might confuse those more familiar with the braneworld literature. In the braneworld literature, a normalizable mode is one that vanishes far away from the brane.\footnote{More precisely, the linearised equations of motion for the transverse-traceless fluctuations in the braneworld setup involve a self-adjoint operator allowing us to define an inner product for the corresponding modes. A normalizable mode is one whose norm is finite under this inner product.} In the exterior $+$ spacetime, the braneworld definition of normalizable and non-normalizable mode agrees with the holographic definition.  This is because the exterior spacetime contains the conformal boundary. However, in the interior $-$ spacetime, the region far away from the brane includes the centre of AdS as opposed to its boundary. This implies that the braneworld definition of normalizable and non-normalizable modes is flipped relative to the holographic definition. From a physics point of view, this is not a problem -- but it can understandably cause confusion when the two communities come together! 

Still, all can agree that to recover 4d gravity on the bubble, we need a graviton zero mode. The wavefunction for a zero mode scales like the background, growing as $\varrho^{-1}$ as we approach the conformal boundary.  This corresponds to the non-normalizable mode by the holographic definition and must be supported by the source on the boundary. In other words, it is sourced by the boundary energy momentum tensor, $\langle\mathcal{T}_{ij}^+\rangle$, given by Eq. \eqref{onepoint} for $\ell=\ell_+$. This is the essence behind the fix in the dark bubble proposal, but it can also apply to our two-sided case for finite $\ell_+$. However, as we send $\ell_+ \to 0$, we see that the boundary energy momentum tensor vanishes, $\langle\mathcal{T}_{ij}^+\rangle \to 0$. In other words, in this particular limit,  we can support a graviton zero mode in the absence of a boundary energy momentum tensor. Gravity is fully localised on the bubble wall in the $\ell_+ \to 0$ limit \emph{without} any sources on the conformal boundary of the exterior spacetime. This now agrees with the one-sided case, where gravity is localised on the ETW brane without any boundary gymnastics.  

\newpage

\section{Discussion}
In this paper we discussed Bubbles of Something and in particular the nucleation of AdS$_5$ from nothing, using the Coleman-de Luccia prescription. We showed that it is possible to obtain a dS$_4$ universe on the boundary ETW brane. Our proposal was motivated by the Cobordism Conjecture within the Swampland programme, which suggests that such ETW branes are necessary defects in quantum gravity and in the absence of global symmetries. It is therefore interesting to understand whether they can be used for braneworld model building. For example, in order to obtain the correct brane tension, we are forced to nucleate AdS$_5$ in the bulk. This could originate from a well-studied background in string theory, such as $\text{AdS}_5\times S^5$ or $\text{AdS}_5\times S^5\slash \mathbb{Z}_k$ \`a la \cite{Horowitz:2007pr}. A similar line of thought was suggested in \cite{Basile:2020mpt} for the Bubble of Nothing. Although we did not discuss the internal manifold explicitly, our work should serve as a proof of concept to pursue such string embeddings.

In understanding the r\^ole of spacetime in quantum gravity, it becomes significantly meaningful to understand the absence of it. Brown and Dahlen proposed that `nothing' can be interpreted as AdS with $\Lambda\to -\infty$. From a holographic point of view this is not bewildering; the dual theory on the boundary simply does not describe any propagating degrees of freedom. However, for the Bubble of Something, it leads to the impression that the brane requires infinite negative tension in order to satisfy the Israel junction conditions. We argue that this is not entirely accurate, and is instead misled by the fact that the holographic dual description is un-renormalized. We gave a brief overview of holographic renormalization and showed that, indeed, all the associated infinities cancel once it is taken into account. The domain wall then effectively becomes a boundary of the spacetime.

So far we focused purely on the gravitational sector of the theory. From the point of view of both cosmology and fundamental physics, it will be interesting and necessary to study dynamics by including additional ingredients such as bulk scalar fields or gauge fields. These will also need to be accounted for in the holographic renormalization, and one has to check that the Brown-Dahlen picture remains consistent.

Since we started by motivating our work through the Swampland programme and some of its conjectures, allow us to also make some final remarks in this direction. It has become increasingly well-appreciated that ETW branes can be associated to infinite distance limits in moduli space \cite{Angius:2022aeq, Buratti:2021fiv}. I.e. as one approaches the ETW brane in the bulk, the vev of some modulus undergoes a diverging field excursion. The simplest example is of course circle compactification, where the radius of the circle pinches off, $R\to 0$, as the radion traverses infinite distance. Invoking the Swampland Distance Conjecture, there is an associated tower of winding modes becoming light, $m_\text{w}\sim R$, gradually spoiling the EFT. But the Swampland Distance Conjecture also describes a dual limit, namely the decompactification, $R\to \infty$, where instead the KK tower becomes light, $m_\text{KK}\sim 1/R$. Viewing nothing as an AdS vacuum -- albeit a peculiar one -- we can surmise whether there is a similar analogy in terms of the AdS Distance Conjecture \cite{Lust:2019zwm}. The vanilla version of the conjecture states that in taking a flat space limit, $\abs{\Lambda}\to 0$, there is an associated tower of states becoming light, $m_\text{tower}\sim \abs{\Lambda}^\alpha$ with $\alpha\sim \mathcal{O}(1)$, causing the EFT to break down. Does this also possess a ``dual'' statement? That is, as $\abs{\Lambda}\to \infty$, is there always a tower of states becoming light, $m_{\widehat{\text{tower}}}\sim \abs{1/\Lambda}^{\widehat{\alpha}}$, with $\widehat{\alpha}\sim \mathcal{O}(1)$ (and possibly related to $\alpha$)? It is at the very least consistent with the generalized notion of distance between vacua proposed in \cite{Mohseni:2024njl}.\footnote{The distance between two AdS$_d$ vacua is suggested to be (in $d$-dimensional Planck units)
\begin{equation}
    \Delta = \gamma\log(\ell_+/\ell_-)
\end{equation}
with $\gamma$ some constant related to the nature of the tower of states becoming light. Indeed, both tunnelling from Minkowski ($\ell_+\to \infty$) and nothing ($\ell_+\to 0$) correspond to opposite infinite distance limits ${\Delta \to \pm\infty}$ for any finite final $\ell_-$.} The Brown-Dahlen interpretation of nothing seems to suggest some nuance to this statement, although we acknowledge that it is mainly speculation at this point. Just like there is an oppositeness between decompactification and `pinching off', is there a likewise relation between `nothing' and Minkowski? We leave this as food for thought.

\section*{Acknowledgements}
We would like to thank Bruno Bento, Ulf Danielsson, Ruth Gregory, Ian Moss and Dani Panizo for useful discussions. AP and PMS acknowledge partial support from the STFC Consolidated Grant nos. ST/V005596/1, ST/T000732/1, and ST/X000672/1.

\newpage
\appendix

\section{Asymmetric Shiromizu-Maeda-Sasaki Equations}\label{sec:appendixA}
Throughout the main parts of this paper, we defaulted to techniques coming from holography and holographic renormalization, in order to extract the theory induced on the brane. Here we show that the same results can also be obtained from a standard braneworld computation. With this, we hope to cement that the three perspectives (one-/two-sided AdS bubble nucleation and RSII) discussed in Section \ref{sec:bubbleofsomething} are really synonymous to one another. The derivation of the effective Einstein field equations on the brane is reminiscent to the Shiromizu-Maeda-Sasaki equations \cite{Shiromizu:1999wj}, but generalized to the asymmetric case. Hence, we will start by reviewing the calculations done in \cite{Gergely:2003pn, Gergely:2004ax}, and finally apply it to our model, taking $\ell_+\to 0$. Consider therefore some 5d line element 
\begin{equation}
    \dd s^2 = \dd r^2 + h_{ij}(r,\xi)\dd \xi^i \dd \xi^j
\end{equation}
and a brane at some $r=r_0(\xi)$ that has unit normal $n^M$. In full generality this brane can be in motion and has acceleration $a^M = n^N \nabla_N n^M$. The induced metric and extrinsic curvature are given by the usual
\begin{equation}
    h_{ij} = g_{ij}-n_in_j\,,\qquad K_{ij}=\nabla_i n_j=\frac{1}{2}\mathcal{L}_ng_{ij}\,.
\end{equation}
Following \cite{Shiromizu:1999wj}, one finds from the Gauss-Codazzi equations
\begin{align}
    G_{ij}^{(4)} &= \frac{2}{3}G^{(5)}_{\{ij\}} + \frac{1}{2}G_{k\ell}^{(5)}n^kn^\ell h_{ij} + F_{ij}-\frac{1}{2}Fh_{ij}-E_{ij} \label{eq:appendixGaussCodazzi}\\
    \nabla_j K^j_i - \nabla_iK &= h^j_in^k G^{(5)}_{jk}\\
    R^{(4)} &= R^{(5)} + F - 2J
\end{align}
where we have defined the quantities
\begin{align}
    Q_{\{ij\}} &\equiv \left(h^k_i h^\ell_j - \frac{1}{4}h_{ij}h^{k\ell}\right) Q_{k\ell}\,,\\
    F_{ij} &\equiv KK_{ij} - K_{ik}K^k_j\,,\\
    E_{ij} &\equiv J_{\{ij\}} - \frac{1}{3}G^{(5)}_{\{ij\}}\,, \\
    J_{ij} &\equiv K_{ik}K^k_j - \mathcal{L}_nK_{ij} + \nabla_j a_i - a_i a_j\,.
\end{align}
Bracketed indices thus denote the transverse-traceless projection with respect to $h_{ij}$. Moreover, it is essentially $J_{ij}$ that carries information about the motion of the brane. Next one can impose the Israel junction conditions
\begin{equation}
    [h_{ij}]^+_-=0\,,\qquad [K_{ij}]^+_- = -\kappa_5^2 \left(S_{ij}-\frac{1}{3}Sh_{ij}\right)\implies [K]^+_-=\frac{1}{3}\kappa_5^2 S
\end{equation}
where $S_{ij}$ is the brane energy momentum tensor and $[Q]_-^+\equiv Q_+-Q_-$ is the jump. At this point we depart from \cite{Shiromizu:1999wj} where they assume $\mathbb{Z}_2$ symmetry and $a^M = 0$. We shall however allow for full generality and focus in particular on when the two sides are different. It will become useful for us to define the $\alpha$-average, which will allow us to compare our computations in the one- and two-sided pictures,
\begin{equation}
    \overline{Q} \equiv (1-\alpha)Q_+ + \alpha Q_-\,,\qquad \alpha=\begin{cases}
        0 & \text{purely outside}\\
        \frac{1}{2} & \text{two-sided}\\
        1 & \text{purely inside}
    \end{cases}\,.
\end{equation}
Now, since $h_{ij}$ must be smooth across the brane, then so must $G^{(4)}_{ij}$. This condition can be consistently written in terms of the $\alpha$-average for the cases we are interested in,
\begin{equation}\label{eq:appendixG4average}
    \overline{G}^{(4)}_{ij} = G^{(4)}_{ij}\,.
\end{equation}
When $S_{ij}=-\sigma h_{ij}$ we have pure tension\footnote{One could otherwise consider matter on the brane $S_{ij}=-\sigma h_{ij}+\tau_{ij}$. In fact, adding $\tau_{ij}$ is the only way to check the effective gravitational coupling on the brane since it is defined through the prefactor that shows up in the Einstein field equation. It is shown in \cite{Shiromizu:1999wj} that $\kappa_4^2=\frac{1}{6}\kappa_5^4\abs{\sigma}$.\label{ft:stressenergy}} and one can recast
\begin{equation}\label{eq:appendixG4}
\begin{split}
    G_{ij}^{(4)} &= \frac{2}{3}\overline{G_{\{ij\}}^{(5)}} + \frac{1}{2}\overline{\left(G^{(5)}_{k\ell}n^kn^\ell\right)} h_{ij} + \overline{F}_{ij} - \frac{1}{2}\overline{F} h_{ij} - \overline{E}_{ij}\\
    &= \frac{2}{3}\overline{G_{\{ij\}}^{(5)}} -\left(\frac{1}{2}\overline{\Lambda}_5+\frac{1}{3}\alpha(1-\alpha)\kappa_5^4\sigma^2+ \frac{1}{4}\left(\overline{K}^2-\overline{K}_{ij}\overline{K}^{ij}\right)\right) h_{ij}\\
    &\qquad\qquad\: + \left(\overline{K}\ \overline{K}_{\{ij\}}-\overline{K}_{\{ik}\overline{K}^k_{j\}}\right) - \overline{E}_{ij}
\end{split}
\end{equation}
In the above we have made use of the algebraic identity 
\begin{equation}
    \overline{AB} = \overline{A}\ \overline{B}+\alpha(1-\alpha)[A]^+_-[B]^+_-
\end{equation}
and the junction condition. Furthermore, define
\begin{align}
    \overline{\mathcal{P}}_{ij} & \equiv \frac{2}{3}\overline{G_{\{ij\}}^{(5)}}\,,\\
    \overline{L}_{ij} &\equiv \overline{K}\ \overline{K}_{ij} - \overline{K}_{ik}\overline{K}^k_j - \frac{1}{2}\left(\overline{K}^2-\overline{K}_{k\ell}\overline{K}^{k\ell}\right) h_{ij}\,.
\end{align}
Note that $\overline{L}_{ij}$ is fully determined by the $\alpha$-averaged extrinsic curvature and diverges in the limit $\ell_+\to 0$ whenever $\alpha\neq 1$. We can then rewrite the extrinsic curvature terms in Eq. \eqref{eq:appendixG4} using the trace and transverse-traceless part of $\overline{L}_{ij}$. What remains on the right-hand side can be grouped by the tensor structure and anything that accompanies a factor of $h_{ij}$ therefore conspires towards an effective 4d cosmological constant. Indeed, if we cast the 4d Einstein field equation as
\begin{equation}
    G^{(4)}_{ij} = -\Lambda_4 h_{ij} + \overline{L}_{\{ij\}} - \overline{E}_{ij}+\overline{\mathcal{P}}_{ij}\,,
\end{equation}
the last three terms are all transverse-traceless and there is an effective 4d cosmological constant
\begin{equation}
    \Lambda_4 = \frac{1}{2}\overline{\Lambda}_5 + \frac{1}{3}\alpha(1-\alpha)\kappa_5^4\sigma^2 - \frac{1}{4}\overline{L}\,.
\end{equation}
Finally, for the $O(5)$ instanton, we can plug in our results from Section \ref{sec:two-sidedAdS}. One finds
\begin{equation}\label{eq:appendixLambda4}
    \Lambda_4 = \frac{1}{2}\overline{\Lambda}_5 + 3\overline{\left(\frac{\chi'}{\chi}\right)^2}\,.
\end{equation}
In the limit of $\ell_+\to 0$ we expect $\Lambda_4$ to remain finite in agreement with the main part of the paper. Still, let us recall that on their own, $\Lambda_5^+ =-6 \ell_+^{-2}$ and $\frac{\chi'}{\chi}\vert_+\sim \ell_+^{-1}$ diverge in this limit. There is however a magical cancellation between the bulk and boundary contributions only when $\Lambda_5^+$ is negative. Expanding Eq. \eqref{eq:appendixLambda4} to significant orders in $\ell_+$ shows
\begin{equation}
    \Lambda_4 =  3(1-\alpha)\left(-\frac{1}{\ell_+^2} + \frac{1}{\ell_+^2} + \frac{1}{\chi_b^2}+\mathcal{O}(\ell_+^2)\right) + 3\alpha\left(-\frac{1}{\ell_-^2}+\left(\frac{\chi'_-(r_-)}{\chi_b}\right)^2\right)\,.
\end{equation}
Now we see that the limit $\ell_+\to 0$ is fully consistent. It gives
\begin{equation}\label{eq:appendixLambda4limit}
    \lim_{\ell_+\to 0}\Lambda_4 = 3\left(\ell_-\sinh\left(\frac{r_0}{\ell_-}\right)\right)^{-2} = 3H_0^2
\end{equation}
where $H_0$ is nothing but the Hubble rate on the bubble, as we saw in Eq. \eqref{eq:desitterslice}. This result holds for any value of $\alpha$, which means that the two-sided picture ($\alpha=\frac{1}{2}$) is exactly the same as if we just considered the inside ($\alpha=1$). The brane cosmological constant approaches $\Lambda_4\to 0^+$ for large bubbles $r_0/\ell_-\to \infty$.

\section{More on Perturbations}\label{sec:appendixB}

This appendix serves as a supplement to our discussion of metric perturbations in Section \ref{sec:perturbations}. The aim is to exemplify in further detail how holographic renormalization intertwines with the normalizable and non-normalizable graviton zero modes. Or rather, for the cautious reader, we will perform part of the boundary gymnastics that we posed was superfluous, and demonstrate that one indeed reaches the same conclusion as with our main argument. To recap, we argued that one can support a graviton zero mode on the bubble even in the absence of a boundary energy momentum tensor. In fact, in the limit $\ell_+\to 0$, metric fluctuations from outside the bubble will not influence the effective description at all. For the purpose of showing this, it becomes more instructive to recast the renormalized bubble action in Eq. \eqref{eq:Sbubbleren} as
\begin{equation}
    S_\text{bubble} = S_\text{in}[-] - S_\text{in}[+] + S_\text{bare}[\text{brane}] + S_\text{ren}[+] \,.
\end{equation}
Varying the first three terms with respect to the brane metric $h_{ij}$ gives the usual Israel junction
\begin{equation}\label{eq:barejunction}
    -\frac{2}{\sqrt{-h}}\frac{\delta}{\delta h^{ij}}\left(S_\text{in}[-] - S_\text{in}[+] + S_\text{bare}[\text{brane}]\right) = \frac{1}{\kappa^2}\left[ K_{ij}-K h_{ij} \right]_-^+ - \sigma_\text{bare}h_{ij}\,.
\end{equation}
We already say in Eq. \eqref{eq:expvalTij} how the remaining term, $S_\text{ren}[+]$, relates to the one-point function of some boundary CFT. Allow us to return to this later. First, consider what would happen if we did not include this term when computing the variation of the bubble action. That is, if we do not take into account holographic renormalization at the junction. If Eq. \eqref{eq:barejunction} happened to reduce to the one-sided junction in the limit $\ell_+\to 0$, including metric perturbations, the task would be done. However, it does not. By performing a brute force expansion of Eq. \eqref{eq:Kijpm} in terms of the FG metric, one can compute the outside contribution to the junction. Together with the bare tension, it gives 
\begin{equation}\label{eq:plusjunctionpert}
\begin{split}
    \frac{1}{\kappa^2}\left(K^+_{ij} - K^+h_{ij}\right) - \sigma_\text{bare}h_{ij} = \phantom{-}&\left( -\frac{3}{\kappa^2\ell_+}h_{ij} - \frac{2\ell_+}{\kappa^2}\varrho_+ \hat{g}_{ij}^{(4)} + T_{ij}^\text{junction}[\hat{g}^{(0)}; \varrho_+]\right)\\
    - &\left(-\frac{3}{\kappa^2\ell_+}+\mathcal{O}(\ell_+) + \sigma_\text{ren}\right) h_{ij}\,.
\end{split}
\end{equation}
From this we can infer that the divergent part of the tension cancels, even when taking into account perturbations of $h_{ij}$. But there remains two additional terms -- one that is proportional to $\hat{g}^{(4)}_{ij}$ containing the normalizable zero mode, and another, which is a more complicated function of the boundary metric, 
\begin{equation}\label{eq:Tijjunction}
\begin{split}
    T_{ij}^\text{junction}[\hat{g}^{(0)};\varrho_+] &= -\frac{\ell_+}{\kappa^2}\bigg(  \hat{g}_{ij}^{(2)} -\hat{g}_{ij}^{(0)}\tr \hat{g}^{(2)} - \varrho_+\hat{g}^{(2)}_{ij}\tr \hat{g}^{(2)}\\
    &\qquad\qquad\qquad +\varrho_+\hat{g}_{ij}^{(0)}\tr \left(\hat{g}^{(2)}\right)^2 - 2\varrho_+\hat{g}_{ij}^{(0)}\tr \hat{g}^{(4)} \bigg)\,.
\end{split}
\end{equation}
The way we interpret it is as an effective energy momentum tensor contribution due to the junction. Note that traces here are meant with respect to $\hat{g}^{(0)}_{ij}$ and furthermore that $T_{ij}^\text{junction}$ only depends on $\hat{g}^{(4)}_{ij}$ through its trace. Therefore it is insensitive to normalizable transverse-traceless fluctuations. All in all, whilst the divergent tension is effectively rendered finite, as was also shown in Section \ref{sec:renormalization}, the same cannot yet be said about metric perturbations in the AdS$_5^+$ part of the bulk. However, we should emphasise an important point at this stage: it is useful to think about metric perturbations in the FG expansion because we can immediately associate $\delta\hat{g}_{ij}^{(0)}$ with the non-normalizable zero mode and $\delta\hat{g}_{ij}^{(4)}$ the normalizable zero mode. But the fluctuations that are actually seen by an observer on the bubble are not the FG ones, but rather $\delta h_{ij}=\delta g_{ij}(\varrho_+, \xi)$. In other words, we ought to rescale the FG modes by appropriate conformal factors
\begin{eqnarray}
    \text{Non-normalizable:}\qquad & \delta h_{ij} &= \frac{\ell_+^2}{\varrho_+}\delta \hat{g}_{ij}^{(0)}\,,\\
    \text{Normalizable:}\qquad & \delta h_{ij} &= \ell_+^2\varrho_+\delta \hat{g}_{ij}^{(4)}\,.
\end{eqnarray}
This rescaling instructs how to infer the sensitivity of Eq. \eqref{eq:plusjunctionpert} with respect to the normalizable and non-normalizable modes. For example, the term of the form $\sim \ell_+\varrho_+\hat{g}_{ij}^{(4)}$ appears at first glance harmless in the limit $\ell_+\to 0$ (which accordingly sends $\varrho_+\to 0)$. But the bubble observer would see the normalizable zero mode from the outside as 
\begin{equation}\label{eq:deltahijdivergence}
    -\frac{2}{\kappa^2\ell_+}\delta h_{ij}\,,
\end{equation}
which diverges as $\ell_+\to 0$. How can we save the bubble dweller from these divergent fluctuations? The missing ingredient is indeed $S_\text{ren}[+]$, which we interpret as an ``ambient'' renormalized AdS$_5^+$ action, that contains information about the conformal boundary at infinity and the necessary holographic counterterms thereof. Varying with respect to $h_{ij}$ one obtains
\begin{equation}\label{eq:Srenplus}
\begin{split}
    -\frac{2}{\sqrt{-h}}\frac{\delta S_\text{ren}[+]}{\delta h^{ij}} &= \left(\frac{\varrho_+}{\ell^2_+}\right)\left(-\frac{2}{\sqrt{-\hat{g}^{(0)}}}\frac{\delta S_\text{ren}[+]}{\delta \hat{g}^{(0)ij}} + \mathcal{O}(\varrho_+)\right)\\
    &= \left(\frac{\varrho_+}{\ell^2_+}\right)\left(\langle\mathcal{T}_{ij}^+\rangle + \mathcal{O}(\varrho_+)\right)\,.
\end{split}
\end{equation}
To leading order it is proportional to the one-point function of the energy momentum in the theory dual to AdS$_5^+$. The one-point function is (c.f. Eq. \eqref{onepoint})
\begin{equation}
    \langle\mathcal{T}_{ij}^+\rangle = \frac{2\ell_+^3}{\kappa^2}\hat{g}^{(4)}_{ij} + X_{ij}[\hat{g}^{(0)}]\,,
\end{equation}
where
\begin{equation}\label{eq:Xsource}
    X_{ij}[\hat{g}^{(0)}] = -\frac{\ell_+^3}{\kappa^2}\left(\frac{1}{4}\hat{g}^{(0)}_{ij} (\tr \hat{g}^{(2)})^2 - \hat{g}^{(0)}_{ij}\tr \hat{g}^{(4)} + \left(\hat{g}^{(2)}\right)^2_{ij} - \frac{1}{2}\hat{g}_{ij}^{(2)}\tr \hat{g}^{(2)}\right)
\end{equation}
is a function of the boundary metric, its covariant derivative, and curvature tensors. If one allows $\hat{g}^{(4)}_{ij}$ to fluctuate, these modes grow by a factor $\varrho_+/\ell_+^2$ as they propagate towards the bubble, as can be seen from Eq. \eqref{eq:Srenplus}. In terms of what the bubble observer sees,
\begin{equation}\label{eq:onepointfunction}
    \frac{\varrho_+}{\ell_+^2} \langle\mathcal{T}^+_{ij}\rangle \sim \frac{2}{\kappa^2\ell_+}\delta h_{ij}\,. 
\end{equation}
Just as we had expected, this is precisely the term needed to kill the divergent perturbation in Eq. \eqref{eq:deltahijdivergence}. Combining Eqs. \eqref{eq:barejunction} to \eqref{eq:onepointfunction} the variation of the renormalized bubble action then reads
\begin{equation}\label{eq:Sbubblerenvariation}
    -\frac{2}{\sqrt{-h}}\frac{\delta S_\text{bubble}}{\delta h^{ij}} = -\frac{1}{\kappa^2}\left(K_{ij}^- -K^-h_{ij}\right) - \sigma_\text{ren}h_{ij} + T_{ij}^\text{junction}[\hat{g}^{(0)};\varrho_+] + \frac{\varrho_+}{\ell_+^2}X_{ij}[\hat{g}^{(0)}]\,.
\end{equation}
All that is left to consider is the non-normalizable mode. In general, however, the two remaining source terms do not cancel. We can compute it explicitly using Eq. \eqref{eq:schouten}, yielding
\begin{equation}
    T_{ij}^\text{junction} + \frac{\varrho_+}{\ell_+^2}X_{ij} = \frac{\ell_+}{2\kappa^2}\left(\hat{G}_{ij}[\hat{g}^{(0)}] + \varrho_+ \hat{\mathcal{R}}_{ij}^2[\hat{g}^{(0)}] \right)\,. 
\end{equation}
Here $\hat{G}_{ij}[\hat{g}^{(0)}]$ is the Einstein tensor with respect to $\hat{g}^{(0)}_{ij}$. At subleading order appear terms quadratic in curvature, which we allow ourselves to abbreviate by some tensor $\hat{\mathcal{R}}_{ij}^2$. Just as before, we ought to translate to the language of the bubble observer by a conformal rescaling. However, to leading order the Einstein does not scale under conformal transformations, and so we find
\begin{equation}
    T_{ij}^\text{junction} + \frac{\varrho_+}{\ell_+^2}X_{ij} = \frac{\ell_+}{2\kappa^2}\left(G_{ij}[h] + \ell_+^2\mathcal{R}^2_{ij}[h] + \mathcal{O}(\ell_+^2H_0^{-1})\right)\,.
\end{equation}
For $\ell_+>0$, there appears an induced higher-curvature gravity theory imprinted on the bubble from the outside AdS$_5^+$. Yet, here comes the up-shot: because the prefactor vanishes as we take $\ell_+\to 0$, the non-normalizable zero mode becomes entirely negligible and completely decouples from physics on the bubble. In other words, they are neither needed to recover the one-sided bubble of something, nor do they play an important r\^ole in obtaining localised gravity on the bubble. This is inherently related to the intuition we presented in Section \ref{sec:perturbations}; the boundary energy momentum tensor vanishes, $\langle \mathcal{T}_{ij}^+\rangle \to 0$, so we can support a graviton zero mode even in the absence of any boundary sources.

\newpage
\bibliographystyle{JHEP}
\bibliography{bibliography}

\end{document}